\documentclass{article}

\PassOptionsToPackage{numbers}{natbib}

\usepackage[preprint]{neurips_2025}

\usepackage[utf8]{inputenc} 
\usepackage[T1]{fontenc}    
\usepackage{hyperref}       
\usepackage{url}            
\usepackage{booktabs}       
\usepackage{amsfonts}       
\usepackage{nicefrac}       
\usepackage{microtype}      
\usepackage{xcolor}         
\usepackage{multirow}
\usepackage{graphicx}
\usepackage{subcaption}
\usepackage[table]{xcolor}
\usepackage{amsmath}
\usepackage{amssymb}
\usepackage{wrapfig}

\title{Rest2Visual: Predicting Visually Evoked fMRI from Resting-State Scans}

\author{%
  Chuyang Zhou \\
  The University of Sydney\\
  \texttt{chuyang.zhou@sydney.edu.au} \\
  \And
  Ziao Ji \\
  The University of Sydney \\
  \texttt{ziji4098@uni.sydney.edu.au} \\
  \AND
  Daochang Liu \\
  The University of Western Australia \\
  \texttt{daochang.liu@uwa.edu.au} \\
  \And
  Dongang Wang \\
  The University of Sydney \\
  \texttt{dongang.wang@sydney.edu.au} \\
  \And
  Chenyu Wang \\
  The University of Sydney \\
  \texttt{chenyu.wang@sydney.edu.au} \\
  \And
  Chang Xu \\
  The University of Sydney \\
  \texttt{c.xu@sydney.edu.au} \\
}

\begin{document}

\maketitle

\begin{abstract}
Understanding how spontaneous brain activity relates to stimulus-driven neural responses is a fundamental challenge in cognitive neuroscience. While task-based functional magnetic resonance imaging (fMRI) captures localized stimulus-evoked brain activation, its acquisition is costly, time-consuming, and difficult to scale across populations. In contrast, resting-state fMRI (rs-fMRI) is task-free and abundant, but lacks direct interpretability. We introduce Rest2Visual, a conditional generative model that predicts visually evoked fMRI (ve-fMRI) from resting-state input and 2D visual stimuli. It follows a volumetric encoder–decoder design, where multiscale 3D features from rs-fMRI are modulated by image embeddings via adaptive normalization, enabling spatially accurate, stimulus-specific activation synthesis. To enable model training, we construct a large-scale triplet dataset from the Natural Scenes Dataset (NSD), aligning each rs-fMRI volume with stimulus images and their corresponding ve-fMRI activation maps. Quantitative evaluation shows that the predicted activations closely match ground truth across standard similarity and representational metrics, and support successful image reconstruction in downstream decoding. Notably, the predicted maps preserve subject-specific structure, demonstrating the model’s capacity to generate individualized functional surrogates. Our results provide compelling evidence that individualized spontaneous neural activity can be transformed into stimulus-aligned representations, opening new avenues for scalable, task-free functional brain modeling.

\end{abstract}

\section{Introduction}
\label{sec-introduction}

Functional magnetic resonance imaging (fMRI) is a central neuroimaging tool for noninvasively mapping human brain function~\cite{ogawa1990brain, logothetis2008we, poldrack2024handbook}, enabling stimulus-driven neural responses to be measured across perceptual, motor, and cognitive domains with high spatial resolution. In particular, visual paradigms have revealed topographically organized representations in the cortex and hierarchical processing across visual areas~\cite{wandell2011imaging, naselaris2011encoding}. However, acquiring visually evoked fMRI (ve-fMRI) requires carefully synchronized experimental protocols, active subject participation, and prolonged scanning time~\cite{dubois2016building, greene2018task}, posing significant barriers to large-scale or clinical deployment~\cite{biswal1995functional, smith2013resting, finn2015functional}.


In contrast, resting-state fMRI (rs-fMRI) offers a task-free and widely available alternative, collected under passive conditions without stimulus delivery or behavioral demands. Despite its accessibility, rs-fMRI signals are difficult to interpret since they reflect complex mixtures of intrinsic neural fluctuations, physiological noise, and latent mental states, with no direct correspondence to external stimuli. As such, inferring functional content from rs-fMRI remains a major challenge~\cite{smith2013resting, finn2015functional, tavor2016task}. Nonetheless, rs-fMRI is known to encode individual-specific and topologically stable connectivity structure~\cite{finn2015functional, kong2019spatial}, raising the possibility that resting-state dynamics may implicitly embed task-relevant priors that could support stimulus-conditioned activation synthesis.

While recent advances have enabled accurate decoding of visual stimuli from evoked brain activity~\cite{naselaris2009bayesian, kay2008identifying, shen2019deep, takagi2023high, scotti2024mindeye2, gong2024neuroclips, mindeye1}, the inverse direction, predicting voxel-level activation from visual input, remains significantly more challenging. Existing encoding models focus exclusively on stimulus-to-activation mapping, neglecting individual brain state variability~\cite{yamins2014performance, naselaris2011encoding}. Moreover, prior approaches that predict task activations from rs-fMRI typically omit visual information and operate at coarse region-level resolution~\cite{tavor2016task, kim2023swift, cole2016activity, ngo2020connectomic}, limiting their capacity to synthesize fine-grained, stimulus-specific responses. These limitations leave open the question: to what extent can spontaneous and stimulus-driven brain states be jointly modeled to reconstruct individualized evoked responses?

We propose a new conditional generative formulation: synthesizing the corresponding stimulus-evoked activation pattern given a resting-state fMRI volume and a visual image stimuli. This formulation enables us to probe the latent correspondence between spontaneous activity and externally driven responses, and to construct individualized ve-fMRI volumes from passive-state input. Beyond expanding the functional utility of rs-fMRI, this approach opens the door to task-free synthesis of interpretable brain activations grounded in natural image semantics.

We instantiate this formulation with \textbf{Rest2Visual}, a conditional generative model that synthesizes voxel-wise, stimulus-evoked activation maps from rs-fMRI and paired 2D images. Rest2Visual adopts a volumetric encoder–decoder architecture built on a 3D convolutional backbone. The encoder extracts multiscale features from resting-state volumes, while the decoder reconstructs spatially aligned ve-fMRI predictions, modulated by visual embeddings via adaptive group normalization. This architecture enables stimulus-specific semantic control over subject-specific intrinsic representations.

To train and evaluate our model, we construct a large-scale triplet dataset based on the Natural Scenes Dataset (NSD)~\cite{allen2022massive}, aligning each rs-fMRI volume with natural image stimuli and their corresponding ve-fMRI activation maps derived via GLM. Experiments demonstrate that Rest2Visual produces spatially accurate, semantically grounded predictions that preserve subject-specific structure and generalize well across diverse stimuli. Together, our formulation and results establish a new pathway for functional brain modeling without requiring task-based acquisition.

\section{Related Work and Background}
\label{sec-background}

\subsection{Functional MRI and Resting-State Dynamics}

Functional MRI (fMRI) noninvasively measures brain activity by detecting blood-oxygen-level-dependent (BOLD) signals. While traditional fMRI studies rely on task-based paradigms to localize stimulus-evoked responses, resting-state fMRI (rs-fMRI) captures spontaneous fluctuations in the absence of external stimuli. These intrinsic dynamics reveal large-scale connectivity structure~\cite{biswal1995functional,fox2005human,smith2013functional,sporns2014contributions}, encode stable inter-individual traits~\cite{finn2015functional,geerligs2015state}, and correlate with task-evoked activations~\cite{tavor2016task,cole2016activity}. Prior work primarily uses rs-fMRI for network-level connectivity analysis or statistical prediction. In contrast, our approach leverages rs-fMRI as a subject-specific functional prior for synthesizing stimulus-conditioned, voxel-level activation patterns.

\subsection{Modeling Visually Evoked Brain Activity}

Voxel-level modeling of visually evoked activity has a long history in computational neuroscience. Early encoding models map hand-engineered features (e.g., Gabor filters, semantic tags) to fMRI responses~\cite{naselaris2011encoding, kay2008identifying}. More recent models use deep neural networks to extract hierarchical visual features that better align with cortical activations~\cite{yamins2014performance, wen2018neural}. The release of the Natural Scenes Dataset (NSD)~\cite{allen2022massive} has enabled large-scale, subject-specific modeling of stimulus-evoked responses. These studies emphasize the sparse and localized nature of ve-fMRI, especially in early visual areas, motivating architectures capable of capturing spatial precision and individual variability.

\subsection{fMRI-to-Image Decoding and Generative Models}

A parallel research line focuses on reconstructing visual stimuli from evoked fMRI. Initial methods employed Bayesian frameworks~\cite{naselaris2009bayesian}, while recent works leverage latent diffusion models conditioned on brain activity~\cite{takagi2023high, mindeye1, scotti2024mindeye2, chen2023seeing}. These models typically assume access to task fMRI and restrict inputs to visual cortex regions (e.g., V1–V5). While informative for evaluating the semantic content of ve-fMRI, these approaches are not applicable to resting-state settings. In this work, we use image decoding as a downstream probe to assess the representational validity of our synthesized activations.

\subsection{Resting-to-Task Mapping and Conditional Brain Generation}

Recent work explores predicting task activation from rs-fMRI~\cite{tavor2016task, cole2016activity, gonzalez2012whole, cohen2020regression}, revealing systematic correspondence between intrinsic connectivity and task responses. However, these efforts are typically statistical and operate at region-level granularity. More recent neural approaches attempt voxel-wise prediction from resting input~\cite{kim2023swift}, but lack stimulus specificity. Our work introduces a conditional generative formulation that integrates rs-fMRI with visual input to predict stimulus-aligned activations. This enables spatially resolved, subject-specific, and stimulus-conditional synthesis, bridging the gap between spontaneous and task-driven brain states.

\section{Methods}
\label{sec-methods}

\subsection{Dataset Construction}
We construct our dataset from the Natural Scenes Dataset (NSD)~\cite{allen2022massive}, which provides high-resolution 7T fMRI responses from 8 healthy adults viewing thousands of natural scene images sourced from MS COCO~\cite{lin2014microsoft}. Let $\mathcal{S} = \{s_1, \dots, s_8\}$ denote the set of human participants. Subjects \(s_{1}\) and \(s_{5}\) each completed 18 sessions (indexed by \(t=1,\dots,18\)) containing a resting‐state run, while the remaining six participants each completed 10 sessions (indexed by \(t=1,\dots,10\)). For each subject-session pair $(s, t)$, we extract a resting-state fMRI scan $\mathbf{x_{raw}}^{(s,t)} \in \mathbb{R}^{T \times D_s \times H_s \times W_s}$ (with $T$ time points, and $D_s$, $H_s$, $W_s$ denoting the depth, height and width of the 3D brain volume for subject $s$, respectively), and a set of stimulus-evoked fMRI beta volumes $\{\mathbf{y_{raw}}^{(s,t)}_i\}_{i=1}^{750}$ paired with visual stimuli $\{\mathbf{v}^{(s,t)}_i\}_{i=1}^{750}$, where $\mathbf{y_{raw}}_i \in \mathbb{R}^{D_s \times H_s \times W_s}$ and $\mathbf{v}_i \in \mathbb{R}^{3 \times 425 \times 425}$.

\paragraph{Spatial Registration.}
To ensure spatial alignment across subjects, all volumes are registered to MNI152 space~\cite{fonov2009unbiased} via continuous interpolation and resampled to 2\,mm isotropic resolution. The spatially normalized versions are denoted as $\mathbf{x}^{(s,t)}, \mathbf{y}^{(s,t)}_{i}\in\mathbb{R}^{D\times H\times W}$, 
\begin{equation}
\mathbf{x}^{(s,t)} \leftarrow \mathcal{T}_{\text{MNI}}\left(\mathbf{x_{raw}}^{(s,t)}\right), \quad
\mathbf{y}_i^{(s,t)} \leftarrow \mathcal{T}_{\text{MNI}}\left(\mathbf{y_{raw}}_i^{(s,t)}\right).
\end{equation}

\paragraph{Resting-state processing.}
To obtain a subject-specific resting-state connectivity prior that captures intrinsic brain organization, we compute a seed-based GLM map from the posterior cingulate cortex (PCC), a key hub in the default mode network. We define $\Omega_{\text{brain}} \subset \{1, \dots, D\} \times \{1, \dots, H\} \times \{1, \dots, W\}$ which denotes the set of voxels within the brain, as defined by a  MNI152 whole-brain mask. Let $\mathcal{R}_{\text{PCC}} \subset \Omega_{\text{brain}}$ denote the PCC region. For each session $t$, we compute the PCC mean time series $\mathbf{r}^{(s,t)}(\tau)$, where $\tau$ indexes fMRI time points, by averaging the BOLD signal across all voxels in the PCC region:
\begin{equation}
\mathbf{r}^{(s,t)}(\tau) = \frac{1}{|\mathcal{R}_{\text{PCC}}|} \sum_{v \in \mathcal{R}_{\text{PCC}}} \mathbf{x}^{(s,t)}(\tau, v), \quad \tau = 1, \dots, T.
\end{equation}
The denoised PCC trace $\tilde{\mathbf{r}}^{(s,t)}$ is obtained by temporal bandpass filtering and convolving with the canonical hemodynamic response function (HRF). A voxel-wise GLM is then fitted:
\begin{equation}
\mathbf{x}^{(s,t)}(\cdot, v) = \beta^{(s,t)}(v) \cdot \tilde{\mathbf{r}}^{(s,t)} + \varepsilon(v), \quad \forall v \in \Omega_{\text{brain}},
\end{equation}
where $\beta^{(s,t)}(v)$ is the regression coefficient for voxel $v$. We z-score the coefficient as
$\mathbf{z}^{(s,t)}(v) = \frac{\beta^{(s,t)}(v) - \mu^{(s,t)}}{\sigma^{(s,t)}}$, where $\mu^{(s,t)}$ and $\sigma^{(s,t)}$ denote the mean and standard deviation of $\{\beta^{(s,t)}(v)\}_{v \in \Omega_{\text{brain}}}$. The resulting map $\mathbf{z}^{(s,t)} \in \mathbb{R}^{D \times H \times W}$ serves as the resting-state intrinsic function map.

\paragraph{Stimulus-evoked data.}
Each stimulus image $\mathbf{v}^{(s,t)}_i$ is associated with a beta activation map $\mathbf{y}^{(s,t)}_i$ derived via trial-level GLM. To eliminate the extreme outliers introduced by NSD's liberal masking policy, we apply the MNI152 whole-brain mask as follows $\tilde{\mathbf{y}}^{(s,t)}_i(v) = \mathbf{y}^{(s,t)}_i(v)\mathbf1_{\Omega_\text{brain}}(v)$.

\paragraph{Final dataset.}
The full data pipeline can be described as a sequence of transformations, and the final dataset is organized into triplets:
\begin{equation}
\mathcal{D} = \mathcal{S}_{\text{HDF5}} \circ \mathcal{C}_{\text{mask}} \circ \mathcal{R}_{\text{conn}} \circ \mathcal{T}_{\text{MNI}}(\mathbf{x}_{\text{raw}}, \mathbf{y}_{\text{raw}}, \mathbf{v}) = \left\{ \left( \mathbf{z}^{(s,t)}, \mathbf{v}^{(s,t)}_i, \tilde{\mathbf{y}}^{(s,t)}_i \right) \mid s \in \mathcal{S}, \, t , \, i\right\},
\end{equation}

where $\mathcal{T}_{\text{MNI}}$ is spatial normalization, $\mathcal{R}_{\text{conn}}$ computes PCC-seeded connectivity, $\mathcal{C}_{\text{mask}}$ applies the brain mask, and $\mathcal{S}_{\text{HDF5}}$ denotes structured storage.

\subsection{Rest2Visual}

Given a subject-session input pair \((s,t)\) and a stimulus index \(i\), the Rest2Visual model takes as input a resting-state connectivity map \(\mathbf{z}^{(s,t)}\) and a corresponding visual stimulus \(\mathbf{v}_i^{(s,t)} \), and predicts the associated stimulus-evoked activation map \(\hat{\mathbf{y}}_i^{(s,t)}\). The model uses a standard conditional encoder–decoder topology based on 3D convolutions.

We denote the full model as a function \(f_\theta: (\mathbf{z}, \mathbf{v}) \mapsto \hat{\mathbf{y}}\), parameterized by \(\theta\), and decomposed it into an image encoder $\mathcal{E}_{\text{img}}: \mathbb{R}^{3 \times H_v \times W_v} \to \mathbb{R}^{C}$ to extract visual stimuli embedding, a volumetric encoder $\mathcal{E}_{\text{vol}}: \mathbb{R}^{D \times H \times W} \to \mathbb{R}^{C_L \times D' \times H' \times W'}$ to extracts multiscale volumetric features from $\mathbf{x}_{\text{rest}}$, and a conditional decoder $\mathcal{D}_{\text{cond}}: \mathbb{R}^{C_L \times D' \times H' \times W'}, \mathbb{R}^C \to \mathbb{R}^{D \times H \times W}$ to reconstructs spatially aligned output incorporating semantic modulation. The decoder integrates the image embedding at every scale via adaptive group normalization. The overall process can be formulized as:

\vspace{-0.5em}
\begin{equation}
\hat{\mathbf{y}}_i^{(s,t)} = f_\theta\left( \mathbf{z}^{(s,t)}, \mathbf{v}_i^{(s,t)} \right) 
= \mathcal{D}_{\text{cond}}\left( \mathcal{E}_{\text{vol}}(\mathbf{z}^{(s,t)}), \, \mathcal{E}_{\text{img}}(\mathbf{v}_i^{(s,t)}) \right).
\end{equation}
\vspace{-0.5em}


The model is trained to minimize voxel-wise prediction error between the generated activation \(\hat{\mathbf{y}}_i^{(s,t)}\) and the corresponding ground-truth beta volume \(\tilde{\mathbf{y}}_i^{(s,t)}\). We use a standard mean squared error (MSE) loss over all brain voxels:
\begin{equation}
\mathcal{L}_{\text{MSE}} = \frac{1}{|\Omega_{\text{brain}}|} \sum_{v \in \Omega_{\text{brain}}} \left( \hat{\mathbf{y}}_i^{(s,t)}(v) - \tilde{\mathbf{y}}_i^{(s,t)}(v) \right)^2.
\end{equation}


During training, each sample \(\left( \mathbf{z}^{(s,t)}, \mathbf{v}_i^{(s,t)}, \tilde{\mathbf{y}}_i^{(s,t)} \right)\) is independently drawn from the dataset \(\mathcal{D}\), and the loss is averaged over the batch. Model parameters \(\theta\) are optimized to minimize \(\mathcal{L}_{\text{MSE}}\) across the dataset:
\begin{equation}
\theta^{*} = \arg\min_{\theta} \, \mathbb{E}_{(s,t,i) \sim \mathcal{D}} \left[ \mathcal{L}_{\text{MSE}} \left( f_\theta(\mathbf{z}^{(s,t)}, \mathbf{v}_i^{(s,t)}), \tilde{\mathbf{y}}_i^{(s,t)} \right) \right].
\end{equation}

\section{Experiments}
\label{sec-experiments}

\subsection{Dataset}

We train our model in a cross-subject setting, where a single model is trained jointly on data from all 8 participants in the NSD. For evaluation, we construct a test set by uniformly sampling 1,000 subject-stimulus pairs from all subjects, ensuring even coverage across participants and stimuli. These samples are held out entirely from training and used exclusively for testing. The remaining samples are randomly split into 90\% for training and 10\% for validation. All splits include data from multiple sessions and reflect the full diversity of brain states and stimulus conditions.

For region-specific evaluations, we adopt two anatomical masks from Nilearn~\cite{abraham2014machine}. One full-brain mask obtained by downsampling the MNI152 brain mask to 2mm resolution, and one visual cortex mask covering areas V1--V5, derived from the Julich-Brain Cytoarchitectonic Atlas \citep{amunts2020julich}.

\subsection{Baselines}

To interpret the significance of our model’s predictions, we introduce two non-parametric baselines that serve as reference points for assessing structural content and stimulus specificity.

\paragraph{Random baseline.}
This lower-bound baseline generates spatial Gaussian noise with the same shape and distribution as the target fMRI volume and treats it as the predicted output. It contains no task-relevant structure, spatial organization, or biological prior, and approximates the statistical profile of non-responsive voxels in GLM-processed fMRI \citep{friston1994statistical, lindquist2008statistical}. Comparing such unstructured outputs to the ground truth assesses whether our evaluation metrics can discriminate signal from background and whether models inadvertently overfit to spatial sparsity.

\paragraph{Resting baseline.}
This baseline uses the subject’s own resting-state fMRI as prediction, without access to the stimulus or any learned mapping. It probes the extent to which intrinsic functional topology aligns with task-induced responses \citep{smith2013functional, cole2016activity}. If our model merely reproduces this structure, performance would match the resting baseline. Substantial improvements beyond it would indicate successful modulation of resting priors by visual input to yield stimulus-aligned predictions.


\subsection{Evaluation}

We evaluate our model's ability to reconstruct visually-evoked fMRI responses using a combination of voxel-level similarity metrics, representational structure preservation, contrast-based region analysis, and downstream decoding performance. All evaluations are conducted on held-out test samples.

\paragraph{Voxel-wise similarity metrics.}
We assess prediction accuracy using three standard voxel-level metrics: Pearson correlation, mean absolute error (MAE), and structural similarity index (SSIM). All fMRI volumes are first $z$-scored within the evaluation region to remove inter-volume scale variability and ensure comparability. Evaluation is performed over both the full brain volume and the visual cortex. Metrics are averaged across subjects and test stimuli, and reported in Table~\ref{tab:main_result}.


\paragraph{Representational similarity analysis (RSA).}
To assess whether the predicted fMRI volumes preserve the representational structure of stimulus responses, we compute representational dissimilarity matrices (RDMs) by measuring pairwise correlation distance between activation patterns across all samples. RSA is then quantified as the Pearson correlation between the upper triangular entries of the predicted and ground-truth RDMs \citep{kriegeskorte2008representational}. This evaluates the extent to which the model captures inter-stimulus geometry, independent of voxel-wise alignment.

\paragraph{SNR and CNR in visual cortex.}
We further assess the signal concentration and spatial contrast of predicted activations using signal-to-noise ratio (SNR) and contrast-to-noise ratio (CNR) metrics, computed within the visual cortex ROI. SNR is defined as the mean activation within the ROI divided by its standard deviation; CNR measures the activation difference between the ROI and the background brain region, normalized by the ROI variance \citep{welvaert2013definition}. All fMRI volumes are normalized using $z$-scoring over the whole-brain mask. Higher values indicate sharper and more spatially localized activation patterns, and results are summarized in Figure~\ref{fig:snr_cnr_boxplot}.

\paragraph{Perceptual evaluation via visual decoding.}
To assess whether predicted ve-fMRI volumes retain meaningful visual information, we feed both predicted and ground-truth fMRI betas map into the MindEye2 decoder \citep{scotti2024mindeye2} to reconstruct stimuli images. We compare the reconstructions using a suite of similarity metrics across low-level (SSIM, AlexNet\cite{krizhevsky2012imagenet} layers) and high-level (Inception\cite{szegedy2015going}, CLIP\cite{radford2021learning}, SwAV\cite{caron2020unsupervised}) feature spaces. This quantifies the semantic alignment of predicted and ground-truth image reconstructions. Results are reported in Table~\ref{tab:stim_metrics}.

\subsection{Main Quantitative Results}

Table~\ref{tab:main_result} reports reconstruction performance across the full brain and the visual cortex. Our model achieves the highest scores across all voxel-level and representational metrics, indicating its ability to reconstruct stimulus-evoked activation maps with high spatial and structural fidelity.

Performance improvements are especially pronounced in the visual cortex, where stimulus-driven responses are strongest. These results validate our core hypothesis: conditioning resting-state fMRI with visual stimuli enables the model to recover sparse, localized activation patterns that reflect task-evoked neural responses. The superiority over unstructured (Random) and stimulus-agnostic (Resting) baselines further confirms that the model does not merely replicate intrinsic connectivity or background noise, but captures meaningful transformations induced by external visual input.

\begin{table}[t]
\centering
\begin{tabular}{c|ccccc}
\toprule
\textbf{ROI} & \textbf{Method} & \textbf{Pearson Corr (↑)} & \textbf{RSA (↑)} & \textbf{MAE (↓)} & \textbf{SSIM (↑)} \\
\midrule
\multirow{5}{*}{\shortstack{Full \\ Brain}}
  & Random  & 0.0000 $\pm$ 0.0020 & 0.0016 & 1.4932 $\pm$ 0.0269 & 0.6022 $\pm$ 0.0861 \\
  & Resting & 0.0104 $\pm$ 0.0560 & 0.5472 & 2.0633 $\pm$ 0.0741 & 0.4134 $\pm$ 0.0947 \\
   & Ours(ns-stim) & 0.1752 $\pm$ 0.0662 & 0.1054 & 1.4750 $\pm$ 0.0720 & 0.6266 $\pm$ 0.0738 \\
   & Ours(ns-fmri) & 0.3357 $\pm$ 0.1164 & 0.6245 & 1.8644 $\pm$ 0.2538 & 0.4952 $\pm$ 0.1080 \\
  & Ours    & \textbf{0.3861 $\pm$ 0.1238} & \textbf{0.6523} & \textbf{1.2653 $\pm$ 0.1143} & \textbf{0.6935 $\pm$ 0.0727} \\
\midrule
\multirow{5}{*}{\shortstack{Visual \\ Cortex \\ (All)}}
  & Random  & 0.0000 $\pm$ 0.0071 & 0.0024 & 2.1384 $\pm$ 0.2180 & 0.3087 $\pm$ 0.0902 \\
  & Resting & 0.1068 $\pm$ 0.1020 & 0.6864 & 2.1745 $\pm$ 0.2017 & 0.2707 $\pm$ 0.1008 \\
   & Ours(ns-stim) & 0.2788 $\pm$ 0.0961 & 0.0018 & 2.9324 $\pm$ 0.2037 & 0.3209 $\pm$ 0.0724 \\
   & Ours(ns-fmri) & 0.6089 $\pm$ 0.1527 & 0.6864 & 2.1036 $\pm$ 0.3039 & 0.4696 $\pm$ 0.1108 \\
  & Ours    & \textbf{0.6541 $\pm$ 0.1560} & \textbf{0.8199} & \textbf{2.0350 $\pm$ 0.2991} & \textbf{0.5077 $\pm$ 0.1147} \\
\bottomrule
\end{tabular}
\vspace{0.3em}
\caption{Quantitative results on full brain and visual cortex ve-fMRI prediction (mean $\pm$ std). Our model, guided by both image stimulus and resting-state priors, achieves consistently better performance, particularly in visually responsive brain areas.}
\label{tab:main_result}
\vspace{-2em}
\end{table}

\subsection{rs-fMRI Prediction and Visual Stimulus Reconstruction}
We assess the quality of predicted ve-fMRI beyond voxel-level quantitative metrics by analyzing their spatial patterns, signal strength, and ability to support stimulus reconstruction. These evaluations offer insight into whether the model captures meaningful, stimulus-driven neural activity.

\paragraph{Anatomical consistency of predicted activation maps.}

\begin{figure}
    \centering
    \includegraphics[width=\linewidth]{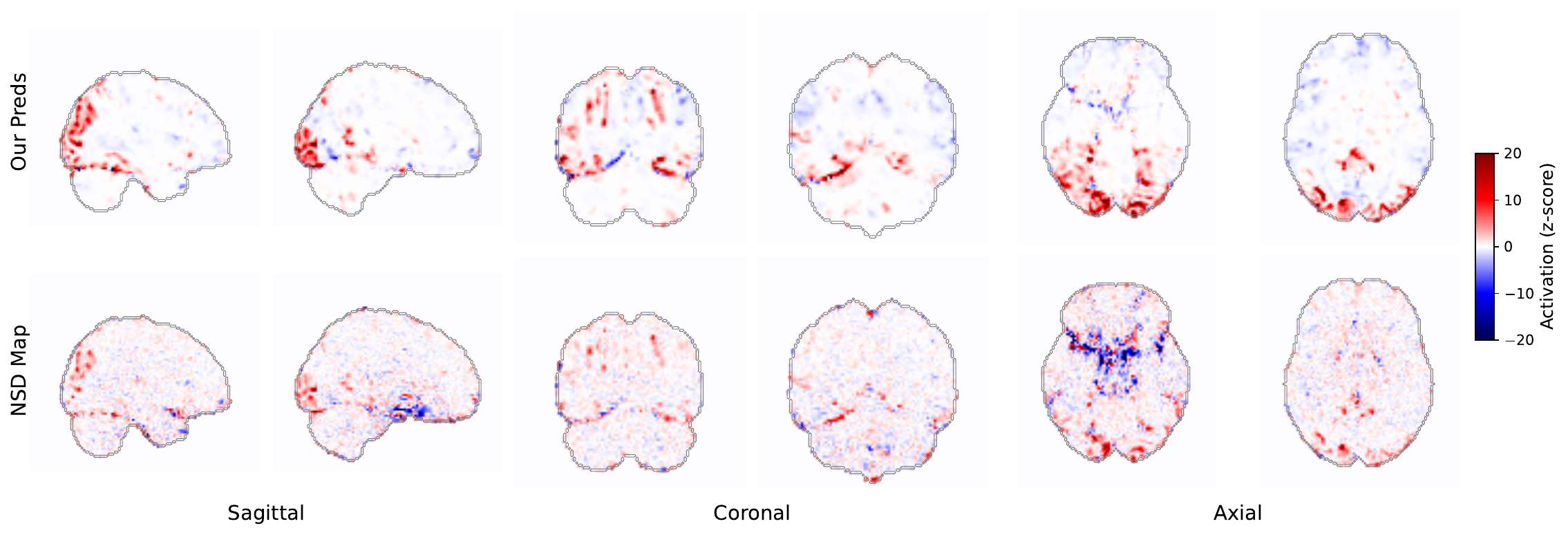}
    \caption{Visual comparison between predicted visually evoked fMRI maps and the NSD-provided ground truth beta maps across multiple anatomical views.}
    \label{fig:multi_view_slices}
    \vspace{-2em}
\end{figure}



\begin{wrapfigure}{r}{0.36\textwidth}
  \centering
  \vspace{-2.5em}
  \resizebox{\linewidth}{!}{\includegraphics[width=\linewidth]{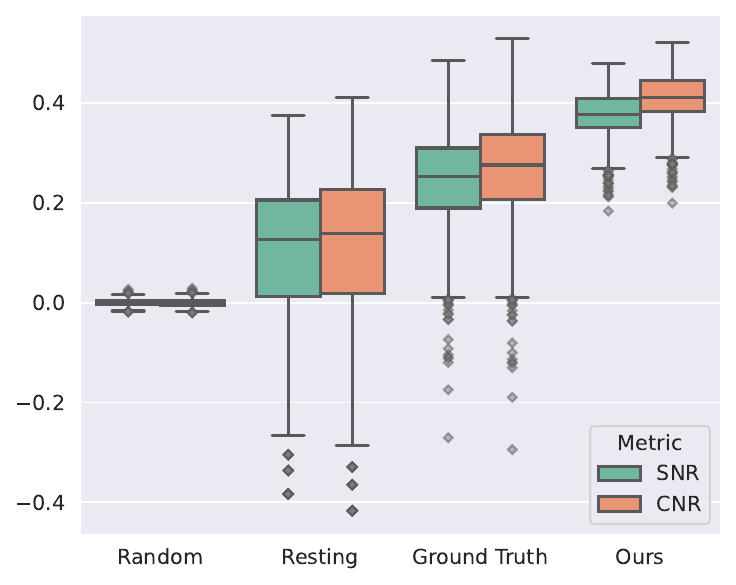}}
  \vspace{-1.5em}
  \caption{Ours demonstrates statistically significant SNR/CNR improvements (two-sided t-test, $p < 1\mathrm{e}{-4}$).}
  \label{fig:snr_cnr_boxplot}
  \vspace{-6.5em}
\end{wrapfigure}

Figure~\ref{fig:multi_view_slices} visualizes predicted ve-fMRI activations alongside ground-truth NSD beta maps. Our predictions exhibit strong alignment with expected retinotopic patterns in early visual areas, including bilateral occipital activations. Compared to the NSD maps, our results are less noisy and demonstrate sharper spatial localization, consistent with the targeted effects of conditioning. The SNR and CNR comparison in Figure~\ref{fig:snr_cnr_boxplot} further confirms our model significantly improves signal clarity and contrast over both baselines and ground truth.

\paragraph{Perceptual integrity of reconstructed images.}

\begin{table}[t]
\centering
\small
\resizebox{1\linewidth}{!}{
\begin{tabular}{c|ccc|ccc}
\toprule
\multirow{3}{*}{\shortstack{\textbf{ve-} \\ \textbf{fMRI}}} & \multicolumn{3}{c|}{Low-Level} & \multicolumn{3}{c}{High-Level} \\
\cmidrule(lr){2-4} \cmidrule(lr){5-7}
 & \textbf{SSIM (↑)} & \textbf{Alex(2) (↑)} & \textbf{Alex(5) (↑)} & \textbf{Incep (↑)} & \textbf{CLIP (↑)} & \textbf{SwAV (↓)} \\
\midrule
NSD  & 0.224 $\pm$ 0.156 & 0.703 $\pm$ 0.110 & 0.555 $\pm$ 0.145 & 0.647 $\pm$ 0.116 & 0.631 $\pm$ 0.126 & 0.444 $\pm$ 0.113 \\
Ours & 0.192 $\pm$ 0.156 & 0.673 $\pm$ 0.115 & 0.512 $\pm$ 0.141 & 0.601 $\pm$ 0.114 & 0.582 $\pm$ 0.122 & 0.496 $\pm$ 0.109 \\
\bottomrule
\end{tabular}
}
\vspace{0.7em}
\caption{Similarity between reconstructed images and original stimuli, using either NSD-provided ve-fMRI or our predicted ve-fMRI as input. Assessing both low-level and high-level feature spaces.}
\vspace{-2em}
\label{tab:stim_metrics}
\end{table}

\begin{figure}[t]
  \centering
  \begin{minipage}[t]{0.111\textwidth}
    \includegraphics[width=\linewidth]{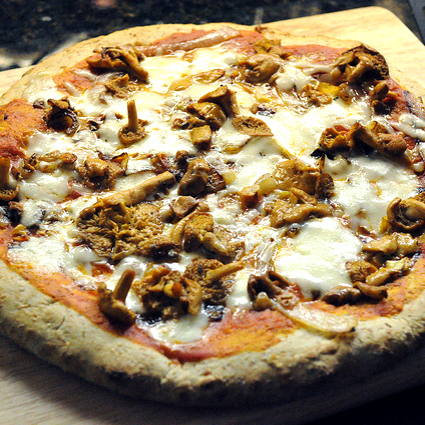}
  \end{minipage}
  \begin{minipage}[t]{0.111\textwidth}
    \includegraphics[width=\linewidth]{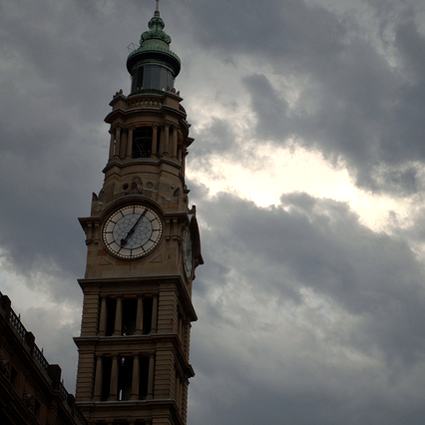}
  \end{minipage}
  \begin{minipage}[t]{0.111\textwidth}
    \includegraphics[width=\linewidth]{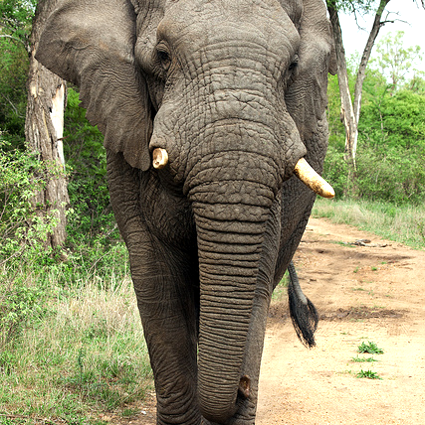}
  \end{minipage}
  \begin{minipage}[t]{0.111\textwidth}
    \includegraphics[width=\linewidth]{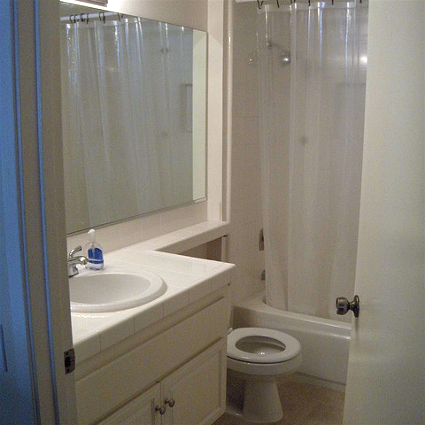}
  \end{minipage}
  \begin{minipage}[t]{0.111\textwidth}
    \includegraphics[width=\linewidth]{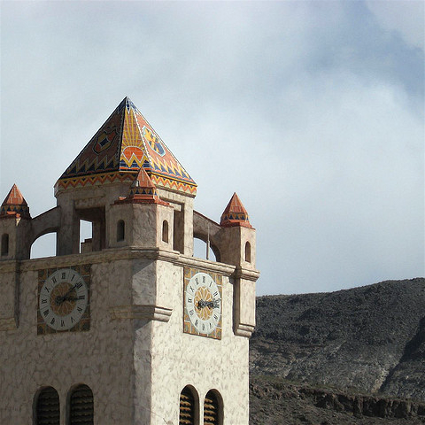}
  \end{minipage}
  \begin{minipage}[t]{0.111\textwidth}
    \includegraphics[width=\linewidth]{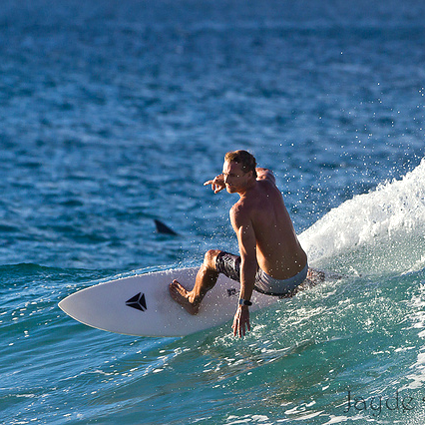}
  \end{minipage}
  \begin{minipage}[t]{0.111\textwidth}
    \includegraphics[width=\linewidth]{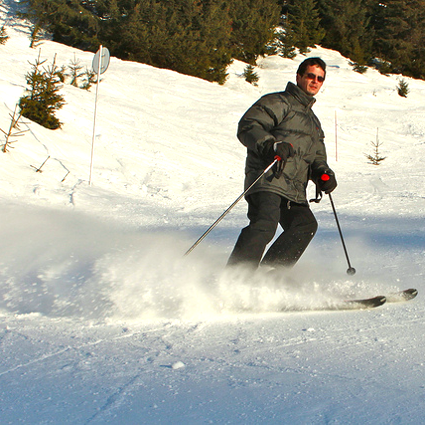}
  \end{minipage}
  \begin{minipage}[t]{0.111\textwidth}
    \includegraphics[width=\linewidth]{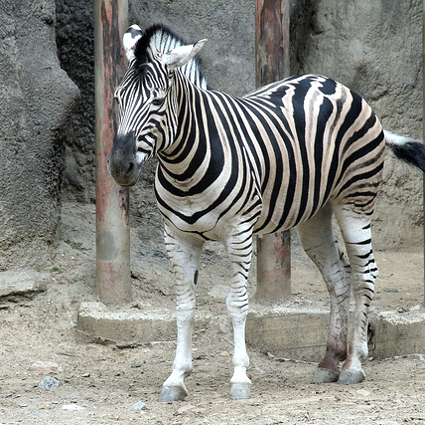}
  \end{minipage}

  \vspace{3pt} 

  \begin{minipage}[t]{0.111\textwidth}
    \includegraphics[width=\linewidth]{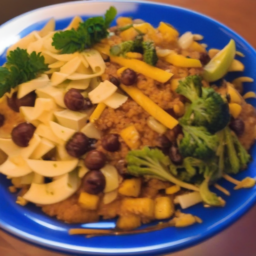}
  \end{minipage}
  \begin{minipage}[t]{0.111\textwidth}
    \includegraphics[width=\linewidth]{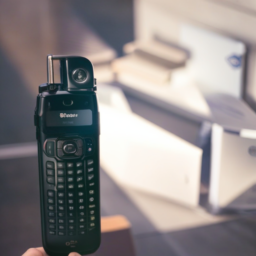}
  \end{minipage}
  \begin{minipage}[t]{0.111\textwidth}
    \includegraphics[width=\linewidth]{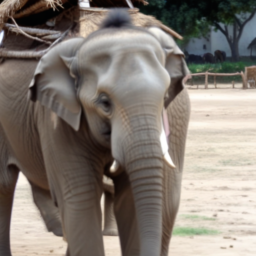}
  \end{minipage}
  \begin{minipage}[t]{0.111\textwidth}
    \includegraphics[width=\linewidth]{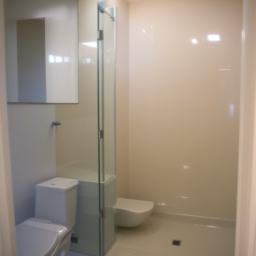}
  \end{minipage}
  \begin{minipage}[t]{0.111\textwidth}
    \includegraphics[width=\linewidth]{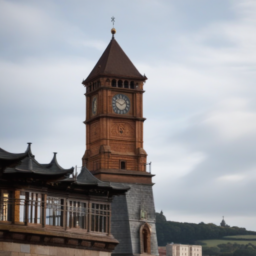}
  \end{minipage}
  \begin{minipage}[t]{0.111\textwidth}
    \includegraphics[width=\linewidth]{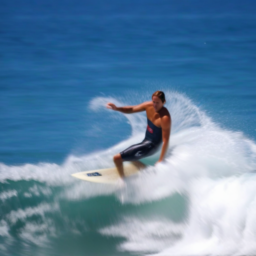}
  \end{minipage}
  \begin{minipage}[t]{0.111\textwidth}
    \includegraphics[width=\linewidth]{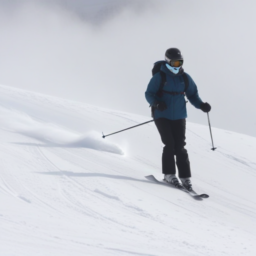}
  \end{minipage}
  \begin{minipage}[t]{0.111\textwidth}
    \includegraphics[width=\linewidth]{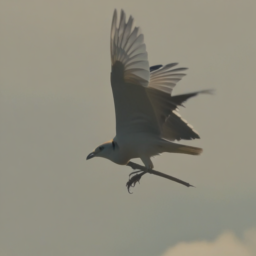}
  \end{minipage}

  \vspace{3pt}

  \begin{minipage}[t]{0.111\textwidth}
    \includegraphics[width=\linewidth]{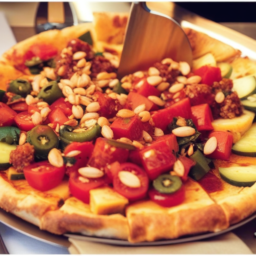}
  \end{minipage}
  \begin{minipage}[t]{0.111\textwidth}
    \includegraphics[width=\linewidth]{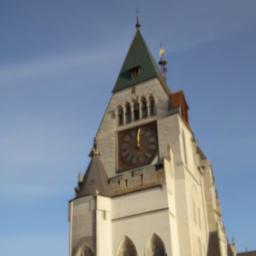}
  \end{minipage}
  \begin{minipage}[t]{0.111\textwidth}
    \includegraphics[width=\linewidth]{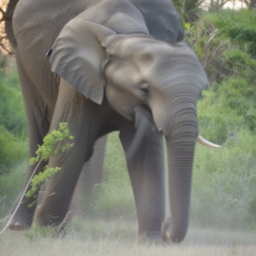}
  \end{minipage}
  \begin{minipage}[t]{0.111\textwidth}
    \includegraphics[width=\linewidth]{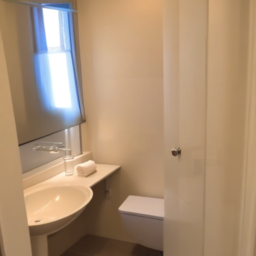}
  \end{minipage}
  \begin{minipage}[t]{0.111\textwidth}
    \includegraphics[width=\linewidth]{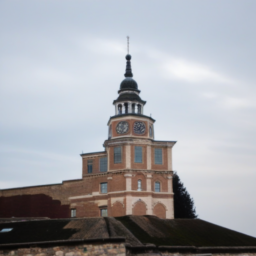}
  \end{minipage}
  \begin{minipage}[t]{0.111\textwidth}
    \includegraphics[width=\linewidth]{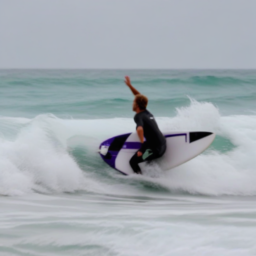}
  \end{minipage}
  \begin{minipage}[t]{0.111\textwidth}
    \includegraphics[width=\linewidth]{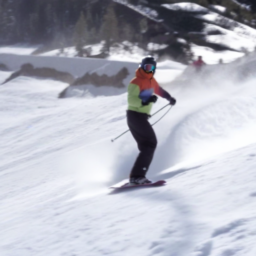}
  \end{minipage}
  \begin{minipage}[t]{0.111\textwidth}
    \includegraphics[width=\linewidth]{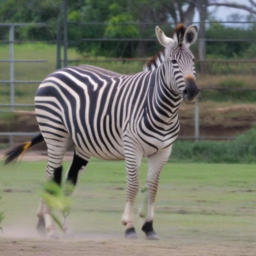}
  \end{minipage}

  \caption{Stimulus images (top row), reconstruction from ground truth ve-fMRI (middle row) and reconstruction from our predicted ve-fMRI (bottom row).}
  \label{fig:recon_demo}
  \vspace{-2em}
\end{figure}

To assess whether predicted ve-fMRI encodes meaningful visual content, we input both predicted and ground-truth fMRI into a pretrained visual decoder (MindEye2) to reconstruct the original stimuli. As shown in Figure~\ref{fig:recon_demo}, both sets of reconstructions retain salient visual properties such as object identity, color, and layout. Table~\ref{tab:stim_metrics} quantifies these similarities across low- and high-level feature spaces. While our predictions score slightly lower than the NSD maps, this discrepancy is expected: MindEye2 is pretrained on NSD beta maps and thus optimized for their specific statistical profile. In contrast, our model produces outputs with stronger stimulus-driven signal but weaker suppression of negative or inhibitory components, which MindEye2 was not trained to decode. Consequently, the slight drop in metrics reflects a domain gap rather than loss of semantic fidelity.

\subsection{Visual Cortex Specific Evaluation}


\begin{wraptable}{r}{0.5\textwidth}
\vspace{-1em}
  \centering
  \resizebox{\linewidth}{!}{\begin{tabular}{cccc}
    \toprule
    \textbf{ROI} & \textbf{Pearson Corr} & \textbf{RSA} & \textbf{SSIM} \\
    \midrule
    V1 & 0.6029 $\pm$ 0.1727 & 0.7077 & 0.4650 $\pm$ 0.1111 \\
    V2 & 0.6243 $\pm$ 0.1662 & 0.7501 & 0.4504 $\pm$ 0.1100 \\
    V3 & 0.7068 $\pm$ 0.1657 & 0.8667 & 0.3845 $\pm$ 0.1422 \\
    V4 & 0.6956 $\pm$ 0.1638 & 0.8635 & 0.4100 $\pm$ 0.1493 \\
    V5 & 0.6345 $\pm$ 0.2075 & 0.7371 & 0.2921 $\pm$ 0.1442 \\
    \bottomrule
  \end{tabular}}
  \vspace{-0.5em}
  \caption{Evaluation across different visual areas.}
  \label{tab:region_eval}
\vspace{-1.5em}
\end{wraptable}

\begin{figure}[t]
\begin{minipage}[b]{0.63\linewidth}

  \centering
  \begin{subfigure}[t]{0.49\textwidth}
    \centering
    \includegraphics[width=\linewidth]{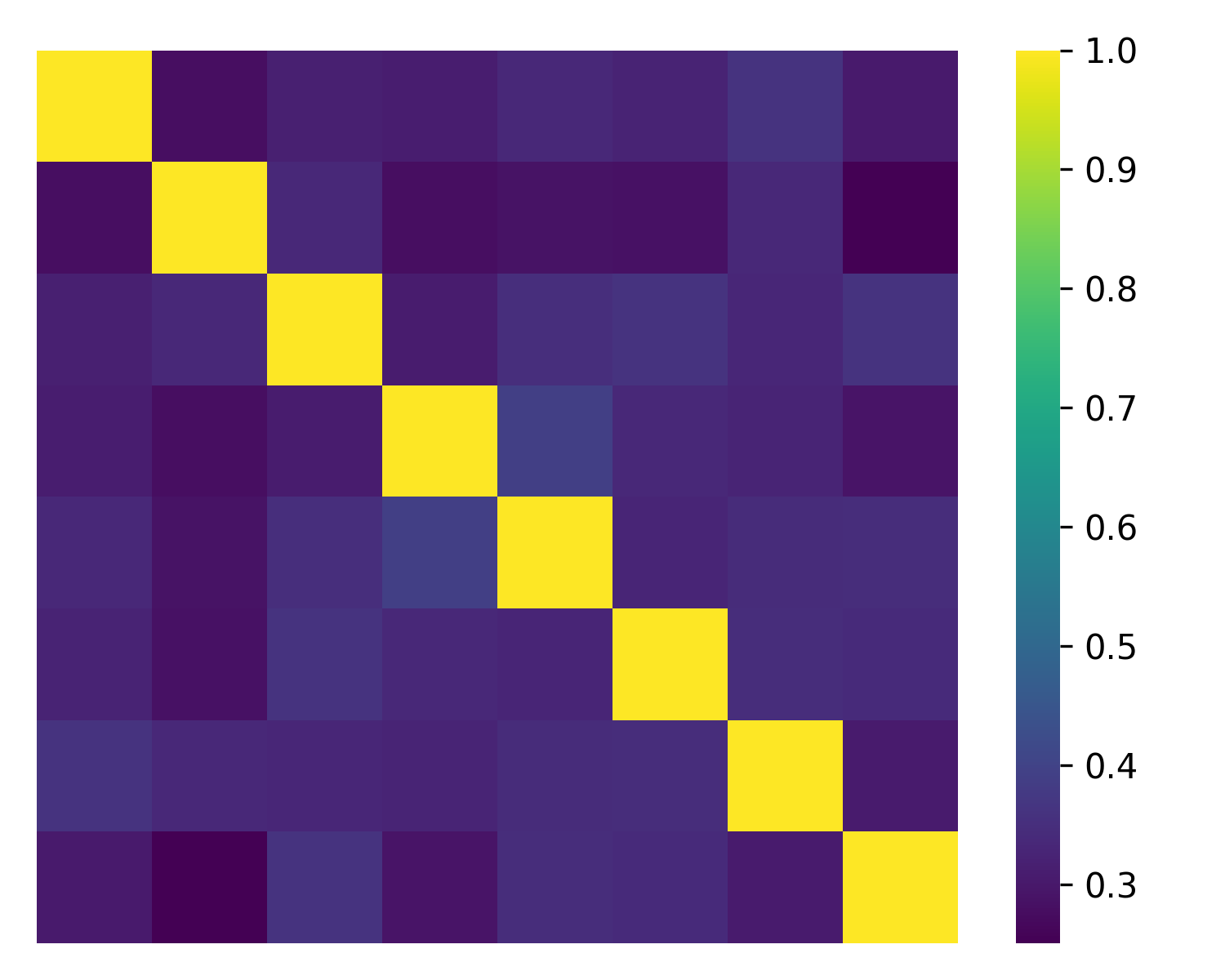}
    \caption{Ours}
    \label{fig:ours_activation}
  \end{subfigure}
  \hfill
  \begin{subfigure}[t]{0.49\textwidth}
    \centering
    \includegraphics[width=\linewidth]{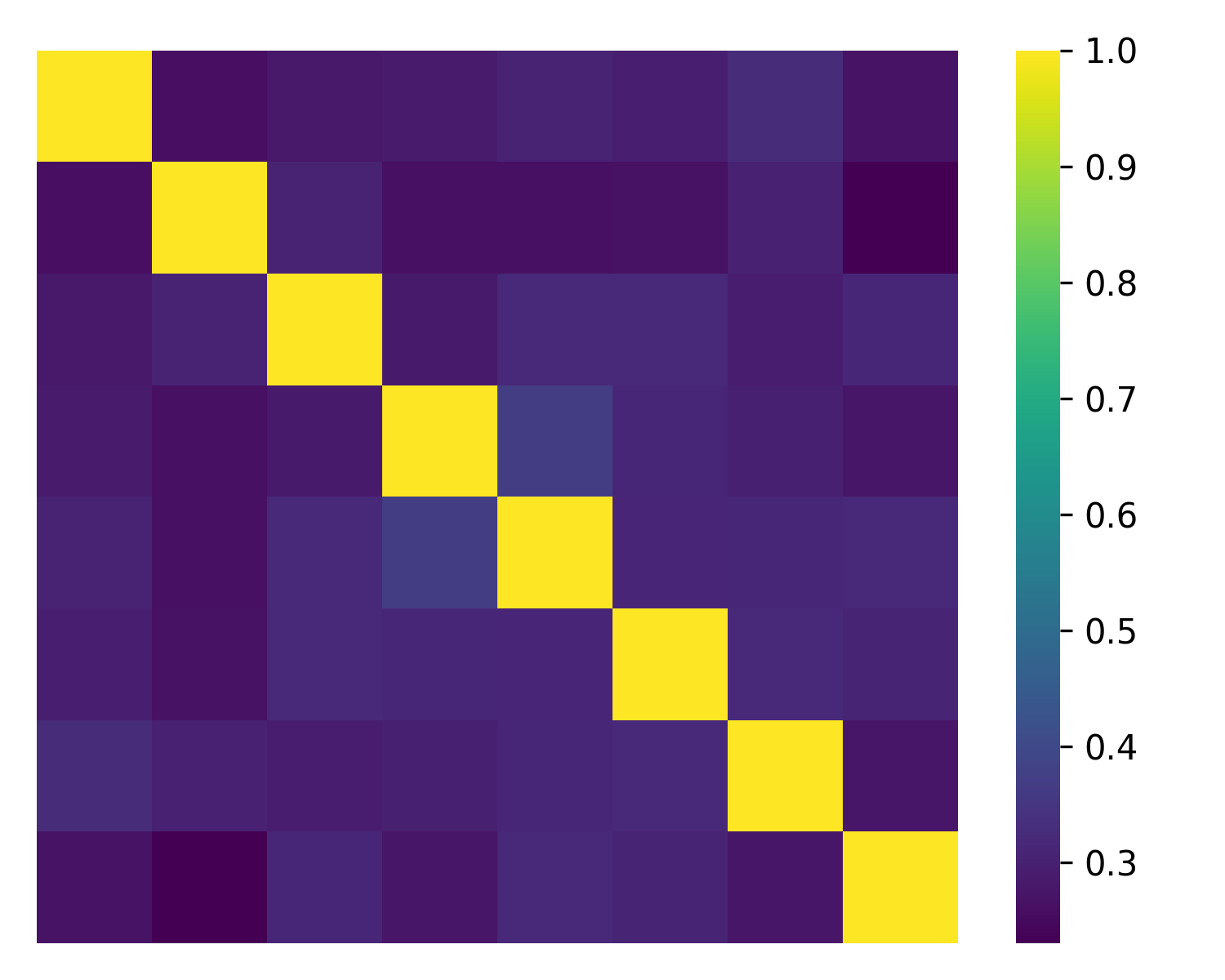}
    \caption{Ground Truth}
    \label{fig:gt_activation}
  \end{subfigure}
  \caption{ Inter-subject correlation matrices (Pearson correlation) of our predicted (a) and ground truth (b) ve-fMRI activation maps, averaged across 125 shared stimuli.}
  \label{fig:activation_comparison}
\end{minipage}
\hfill
\begin{minipage}[b]{0.33\linewidth}
  \centering
  \includegraphics[width=\linewidth]{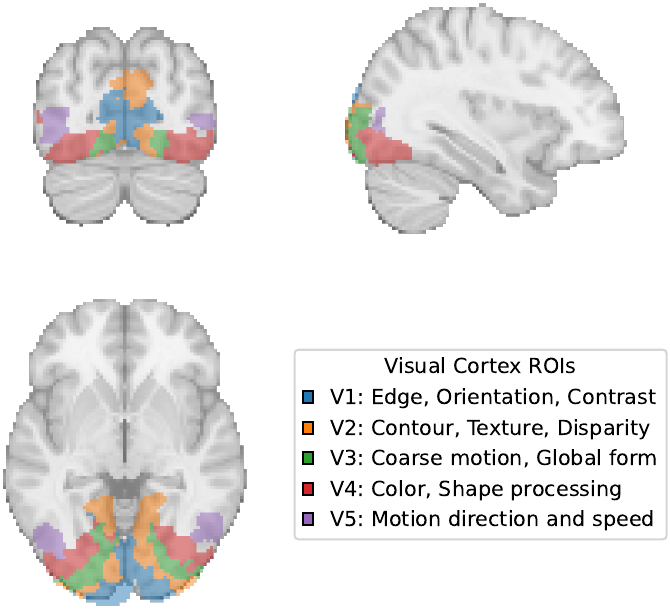}
  \caption{Anatomical distribution of visual cortex ROIs (V1–V5) from the Julich-Brain atlas.}
  \label{fig:region_rois}
\end{minipage}
\vspace{-1em}
\end{figure}

We further analyze the performance across individual visual areas V1–V5, using cytoarchitectonic ROIs from the Julich-Brain atlas~\cite{amunts2020julich} (Figure~\ref{fig:region_rois}), to reveal the region-specific activation patterns captured by the model. These regions span early to mid-level visual processing: V1 encodes local edges and contrast, V2 represents texture and disparity, V3 and V4 are key for shape, color, and semantic content perception, while V5 specializes in motion~\citep{felleman1991distributed, wandell2007visual, hubel1968receptive, zeki1974functional, zeki1980representation}.

As shown in Table~\ref{tab:region_eval}, our model achieves high RSA and Pearson correlation in V3 and V4, reflecting strong preservation of structured stimulus-driven responses. These mid-level areas are closely tied to object perception, suggesting that the model effectively aligns resting-state priors with perceptual demands. While SSIM scores in V3 and V4 are a bit lower, this does not contradict our findings. SSIM emphasizes spatial accuracy, and the more distributed activation patterns in mid-level areas like V3 and V4 can yield lower SSIM even when functional responses are well preserved.

Overall, these results confirm that our model not only captures global brain activation patterns but also aligns with known functional specializations in the visual cortex, exhibiting the strongest effects in regions most critical for interpreting images. This highlights the model’s capacity to bridge intrinsic brain dynamics with stimulus-specific processing in anatomically meaningful ways.

\subsection{Individuality Preservation}

Preserving individual-specific brain responses is central to our modeling objective, yet remains underexplored in existing ve-fMRI prediction frameworks. To assess whether our model maintains subject identity in predicted activation patterns, we evaluate inter-subject correlation matrices over the 125 shared stimuli observed by all participants (Figure~\ref{fig:activation_comparison}). Each matrix entry reflects the Pearson correlation between predicted ve-fMRI responses of two subjects viewing the same set of images.

As shown in Figure~\ref{fig:activation_comparison}, the predicted correlation matrix~\ref{fig:ours_activation} closely resembles the ground truth~\ref{fig:gt_activation}, particularly in its off-diagonal structure, where all values remain low. This structure confirms that our model produces individualized activation patterns that preserve subject-specific functional signatures, rather than collapsing to population-level averages. These findings support a key strength of our approach: by incorporating rs-fMRI as input, the model leverages individual-specific functional architecture to reconstruct personalized stimulus-evoked responses.

\subsection{Ablation Study}

\begin{figure}[t]
  \centering
  \begin{minipage}[t]{0.49\textwidth}
    \centering
    \includegraphics[width=0.24\linewidth]{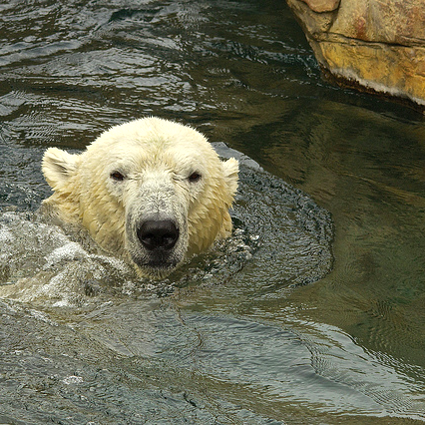}
    \includegraphics[width=0.24\linewidth]{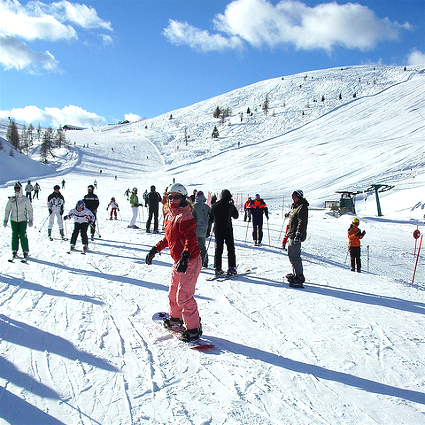}
    \includegraphics[width=0.24\linewidth]{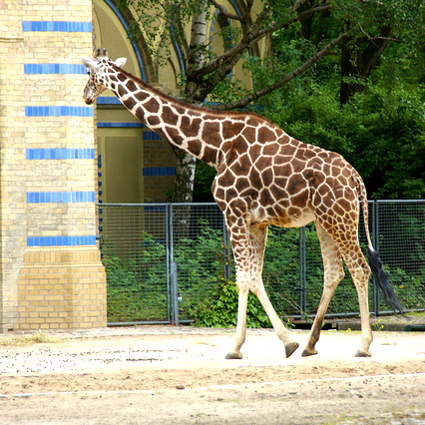}
    \includegraphics[width=0.24\linewidth]{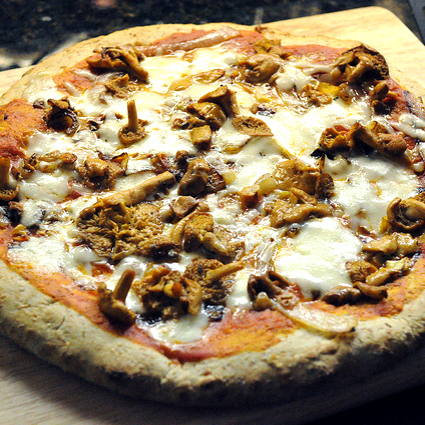}

    \vspace{2pt}

    \includegraphics[width=0.24\linewidth]{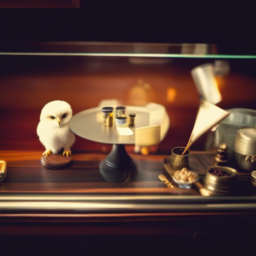}
    \includegraphics[width=0.24\linewidth]{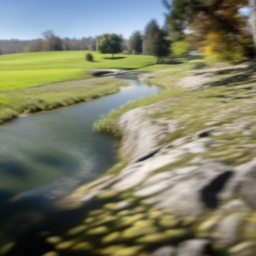}
    \includegraphics[width=0.24\linewidth]{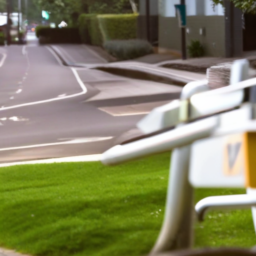}
    \includegraphics[width=0.24\linewidth]{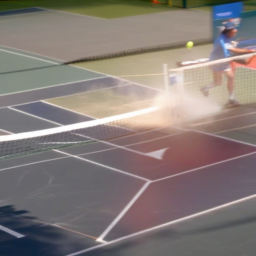}

    \caption*{(a) Noisy resting-state input with valid stimulus}
  \end{minipage}
    \hfill
  \begin{minipage}[t]{0.49\textwidth}
    \centering
    \includegraphics[width=0.24\linewidth]{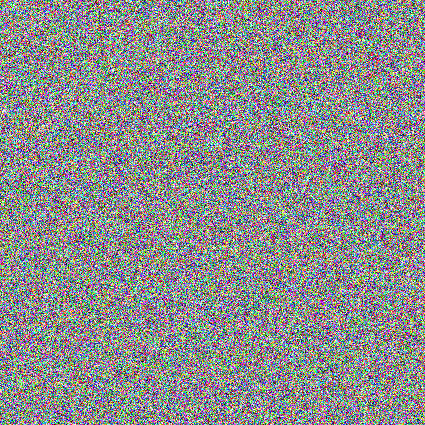}
    \includegraphics[width=0.24\linewidth]{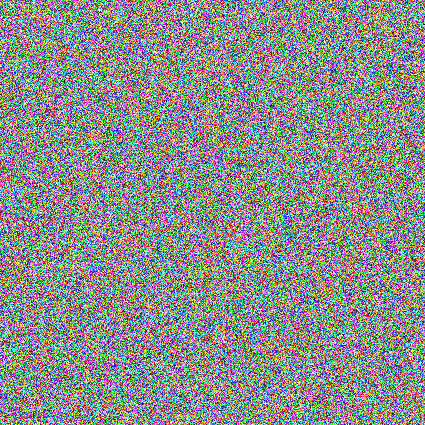}
    \includegraphics[width=0.24\linewidth]{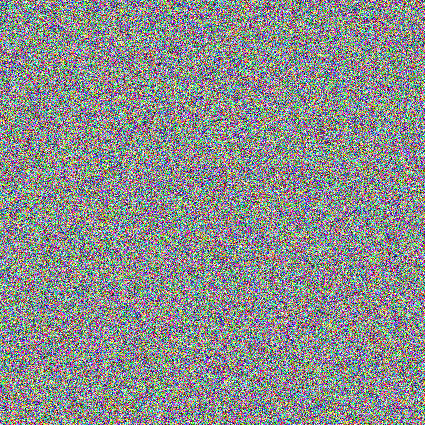}
    \includegraphics[width=0.24\linewidth]{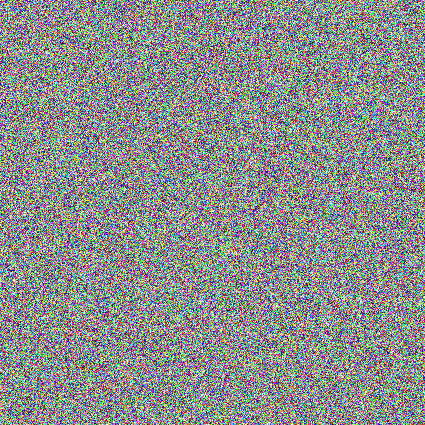}

    \vspace{2pt}

    \includegraphics[width=0.24\linewidth]{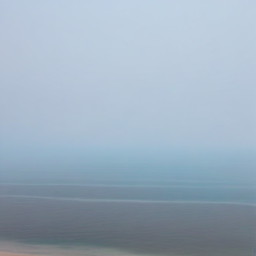}
    \includegraphics[width=0.24\linewidth]{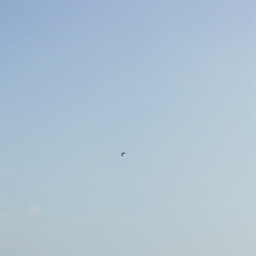}
    \includegraphics[width=0.24\linewidth]{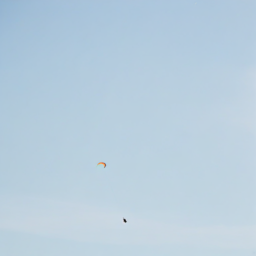}
    \includegraphics[width=0.24\linewidth]{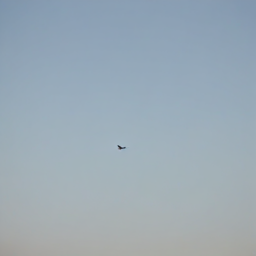}

    \caption*{(b) Noisy stimulus with valid resting-state input}
  \end{minipage}
  \vspace{-0.5em}

  \caption{
    Ablation analysis of input modalities. (a) When resting-state input is replaced with noise, predictions collapse, underscoring the necessity of intrinsic functional structure for meaningful activation. (b) When visual stimuli are replaced with Gaussian noise, reconstructions become visually meaningless but spatially coherent, indicating strong reliance on resting-state priors and confirming that predicted fMRI faithfully encode stimulus information, although semantically meaningless.
  }
  \label{fig:ablation_inputs}
  \vspace{-2em}
\end{figure}

To disentangle the contributions of the resting-state fMRI and visual stimulus inputs, we perform targeted ablations by replacing each modality with Gaussian noise and analyzing the resulting activation maps and corresponding image reconstructions.

\paragraph{Noisy rs-fMRI input (ns-fmri).}
When the resting-state input is replaced by spatial Gaussian noise while preserving the visual stimulus (\textbf{Ours(ns-fmri)}), performance deteriorates sharply across all metrics (Table~\ref{tab:main_result}), and the predicted fMRI patterns become spatially incoherent and biologically implausible. As shown in Figure~\ref{fig:ablation_inputs}, the corresponding image reconstructions are structurally chaotic and semantically meaningless. This is expected since our model learns to exploit subject-specific functional organization from resting-state fMRI to anchor spatial localization, and replacing it with noise entirely removes this inductive prior. The results clearly demonstrate that functional structure from rs-fMRI is indispensable for meaningful reconstruction.

\paragraph{Noisy stimulus input (ns-stimuli).}
In the other setting, we replace the RGB stimulus image with Gaussian noise while keeping the resting-state input intact (\textbf{Ours(ns-stimuli)}). Interestingly, although the input contains no semantic content, the reconstructed fMRI maps retain structured patterns and yield relatively high quantitative scores as shown in Table~\ref{tab:main_result}. As shown in Figure~\ref{fig:ablation_inputs}, the outputs are large homogeneous color blobs devoid of semantic structure which are visually similar to the noisy input image. This suggests that the model has learned to treat the presence of any visual stimulus as a signal to generate corresponding brain activity, even when the stimulus lacks meaningful content.

\paragraph{Interpretation.}
This behavior reveals an important asymmetry: the model treats visual stimulus presence as a strong global modulator of activation patterns, while finer distinctions across stimuli rely on semantic content encoded in the RGB image. The relatively high scores for ns-stimuli should not be misinterpreted as semantic correctness. They arise because most ve-fMRI volumes share global features such as consistent activation in early visual cortex even across different stimuli\cite{kay2008identifying, yamins2014performance, gucclu2015deep}.

\subsection{Generalization Analysis}
\begin{figure}[t]
  \centering
  \begin{minipage}[t]{0.111\textwidth}
    \includegraphics[width=\linewidth]{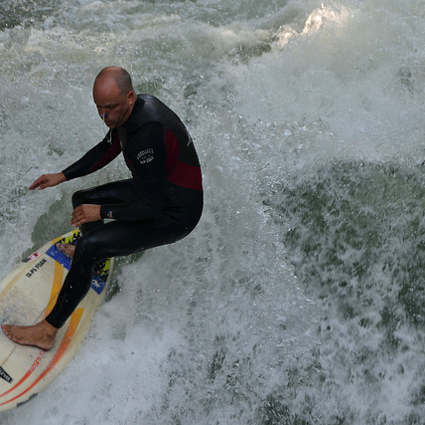}
  \end{minipage}
  \begin{minipage}[t]{0.111\textwidth}
    \includegraphics[width=\linewidth]{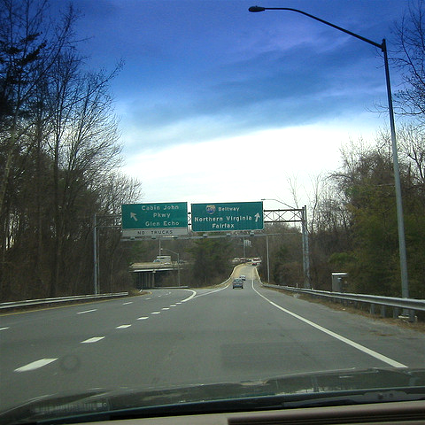}
  \end{minipage}
  \begin{minipage}[t]{0.111\textwidth}
    \includegraphics[width=\linewidth]{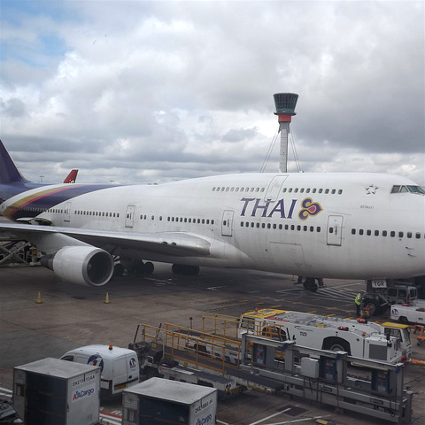}
  \end{minipage}
  \begin{minipage}[t]{0.111\textwidth}
    \includegraphics[width=\linewidth]{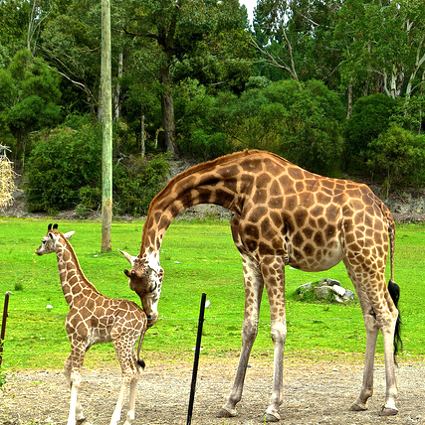}
  \end{minipage}
  \begin{minipage}[t]{0.111\textwidth}
    \includegraphics[width=\linewidth]{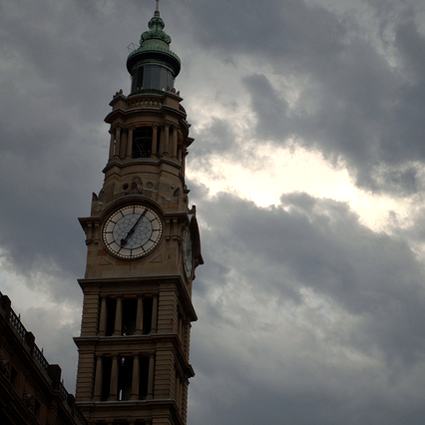}
  \end{minipage}
  \begin{minipage}[t]{0.111\textwidth}
    \includegraphics[width=\linewidth]{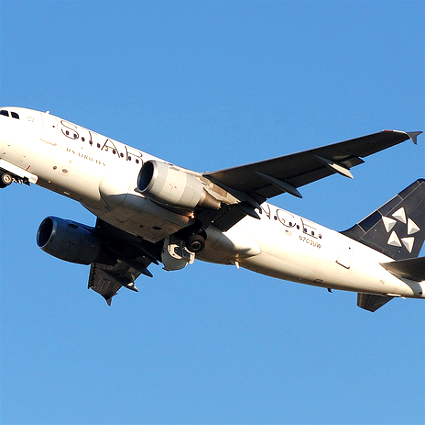}
  \end{minipage}
  \begin{minipage}[t]{0.111\textwidth}
    \includegraphics[width=\linewidth]{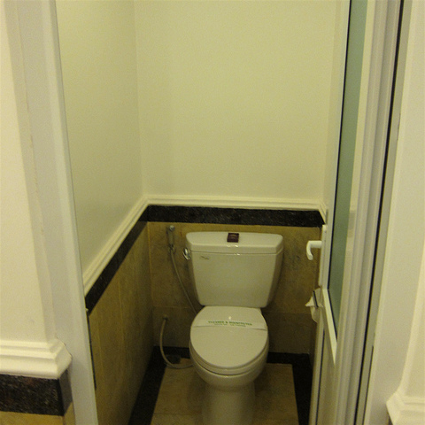}
  \end{minipage}
  \begin{minipage}[t]{0.111\textwidth}
    \includegraphics[width=\linewidth]{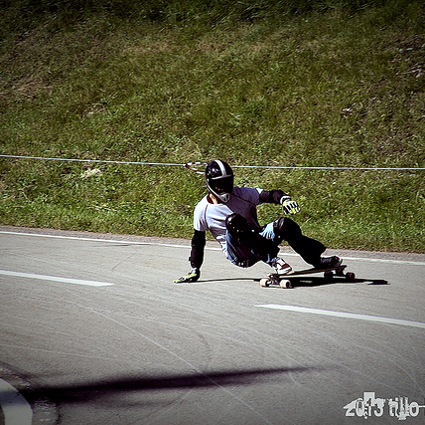}
  \end{minipage}

  \vspace{3pt}



  \begin{minipage}[t]{0.111\textwidth}
    \includegraphics[width=\linewidth]{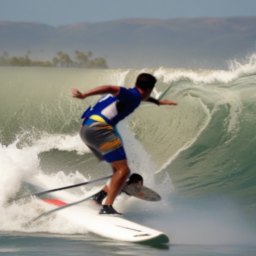}
  \end{minipage}
  \begin{minipage}[t]{0.111\textwidth}
    \includegraphics[width=\linewidth]{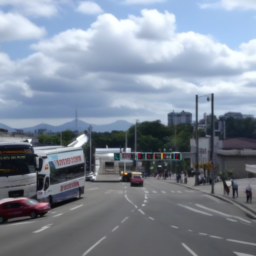}
  \end{minipage}
  \begin{minipage}[t]{0.111\textwidth}
    \includegraphics[width=\linewidth]{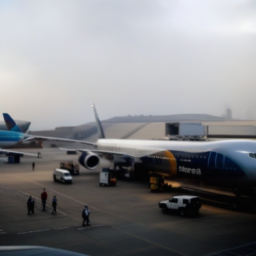}
  \end{minipage}
  \begin{minipage}[t]{0.111\textwidth}
    \includegraphics[width=\linewidth]{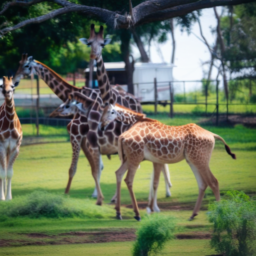}
  \end{minipage}
  \begin{minipage}[t]{0.111\textwidth}
    \includegraphics[width=\linewidth]{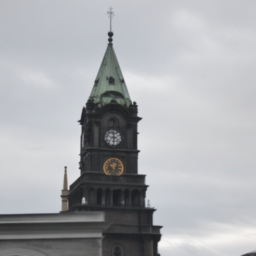}
  \end{minipage}
  \begin{minipage}[t]{0.111\textwidth}
    \includegraphics[width=\linewidth]{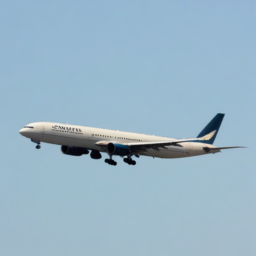}
  \end{minipage}
  \begin{minipage}[t]{0.111\textwidth}
    \includegraphics[width=\linewidth]{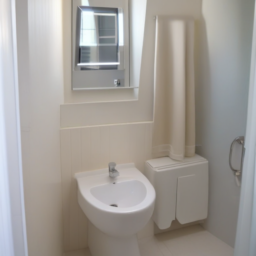}
  \end{minipage}
  \begin{minipage}[t]{0.111\textwidth}
    \includegraphics[width=\linewidth]{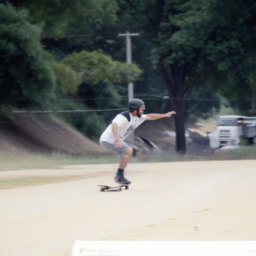}
  \end{minipage}

  \caption{Stimuli images (top) and predicted reconstructions (bottom) for unseen stimulus.}
  \label{fig:unseen}
  \vspace{-1em}
\end{figure}

We evaluate the generalization ability of our model under two challenging settings: unseen visual stimuli and unseen subjects. These settings test whether the model captures transferable principles rather than overfitting to specific images or individuals.

\paragraph{Unseen Stimuli.}
We first assess whether the model can predict reliable activation patterns for novel images never seen during training. As shown in Table~\ref{tab:generalization}, performance remains strong across all metrics. Visual decoding of the predicted ve-fMRI (Figure~\ref{fig:unseen}) yields reconstructions that retain key semantic and structural content. These results suggest that the model learns to map generic visual features, rather than memorized image identities, into corresponding activation patterns. Despite never observing these stimuli before, the model produces brain responses that are both spatially accurate and perceptually aligned with the stimulus content.

\paragraph{Unseen Subject.}
We also assess generalization to a held-out subject by excluding all of their data during training. Performance drops notably, especially in representational similarity and voxel-level alignment, reflecting the model’s dependence on subject-specific resting-state priors. This result is expected given that the NSD dataset includes only eight participants, limiting inter-subject diversity and making cross-subject generalization inherently difficult. The outcome underscores the importance of individual functional architecture in shaping accurate predictions.

\begin{table}[htbp]
\vspace{-0.5em}
\centering
\begin{tabular}{lcccc}
\toprule
\textbf{Method} & \textbf{Pearson} & \textbf{RSA} & \textbf{MAE} & \textbf{SSIM} \\
\midrule
Ours (Normal)  & 0.3861 $\pm$ 0.1238 & 0.6523 & 1.2653 $\pm$ 0.1143 & 0.6935 $\pm$ 0.0727 \\
Ours (Unseen stimuli) & 0.3804 $\pm$ 0.1243 & 0.6438 & 1.2639 $\pm$ 0.1130 & 0.6950 $\pm$ 0.0726 \\
Ours (Unseen subject)   & 0.1361 $\pm$ 0.0430 & 0.2590  & 1.4714 $\pm$ 0.0485 & 0.6992 $\pm$ 0.0377 \\
\bottomrule
\end{tabular}
\vspace{0.5em}
\caption{Generalization evaluation for unseen stimulus and subjects.}
\label{tab:generalization}
\vspace{-2.5em}
\end{table}

\section{Discussion and Future Work}
\label{sec-discussion}

We introduce a new task paradigm for functional brain modeling: predicting visually evoked responses directly from resting-state fMRI and visual input. While most existing models rely solely on either visual stimuli or resting-state data, our approach integrates both, combining task-driven specificity with subject-dependent priors to reconstruct individualized evoked responses. By grounding evoked predictions in passive-state data, our approach enables functional inference without requiring active task participation, which is a property with strong potential for clinical and noncompliant populations.

Our model accurately reconstructs localized, stimulus-driven activation in early visual areas, while also preserving subject-specific response structure and can generalizes to unseen stimuli. These results support the hypothesis that resting-state features can be productively modulated by external input to yield functionally meaningful and individualized predictions.

The primary limitation is generalization across subjects. With only eight participants in the NSD dataset, the model has limited exposure to the full rage of inter-individual variability, which constrains its ability to synthesize accurate predictions for unseen brains. 


Future work may extend this paradigm to time-resolved modeling, incorporate multi-modal inputs such as anatomical or diffusion imaging, or adapt it for developmental, aging, or clinical cohorts. Ultimately, our results highlight the feasibility of decoding evoked brain function from passive-state data, offering a new direction for functional mapping when task-based acquisition is infeasible.


\bibliographystyle{unsrtnat}
\bibliography{ref}

\end{document}